\documentclass[manuscript]{emulateapj-rtx4}

\newcommand{\Ell}{E_\parallel}      
\newcommand{\rhoGJ}{\rho_{{\rm GJ}}}  
\newcommand{\rlc}{\varpi_{\rm LC}} 

\slugcomment{accepted 2011 May}

\shorttitle{Death Line of Gamma-ray Pulsars}
\shortauthors{Wang and Hirotani}

\begin{document}


\title{Death Line of Gamma-ray Pulsars with Outer Gaps}


\author{Ren-Bo Wang\altaffilmark{1}}
\affil{Department of Physics, 
       National Dong Hwa University, 
       Hualien 97401, Taiwan}
\email{rbwang1225@gmail.com}

\and

\author{Kouichi Hirotani\altaffilmark{2}}
\affil{Theoretical Institute for
       Advanced Research in Astrophysics (TIARA),
       Academia Sinica, Institute of Astronomy and Astrophysics (ASIAA),
       PO Box 23-141, Taipei, Taiwan}
\email{hirotani@tiara.sinica.edu.tw}

\altaffiltext{1}{Summer student of ASIAA in 2010}

\altaffiltext{2}{Postal address: 
                 TIARA, Department of Physics, 
                 National Tsing Hua University,
                 101, Sec. 2, Kuang Fu Rd.,Hsinchu, Taiwan 300}




\begin{abstract}
We analytically investigate the condition for a
particle accelerator to be active in the outer magnetosphere
of a rotation-powered pulsar.
Within the accelerator (or the gap),
magnetic-field-aligned electric field
accelerates electrons and positrons, 
which emit copious gamma-rays via curvature process.
If one of the gamma-rays emitted by a single pair
materializes as a new pair on average, the gap is self-sustained.
However, if the neutron-star spin-down rate decreases below
a certain limit,
the gap becomes no longer self-sustained 
and the gamma-ray emission ceases.
We explicitly compute the multiplicity of cascading pairs
and find that the obtained limit corresponds to a modification of 
previously derived outer-gap death line.
In addition to this traditional death line,
we find another death line, 
which becomes important for millisecond pulsars, 
by separately considering the threshold of
photon-photon pair production.
Combining these traditional and new death lines,
we give predictions on the detectability of 
gamma-ray pulsars with Fermi and AGILE.
An implication on the X-ray observations of heated polar-cap
emission is also discussed.
\end{abstract}





\section{Introduction}
\label{sec:intro}
The {\it Fermi} Large Area Telescope (LAT) provides 
a wealth of new data on isolated, rotation-powered pulsars,
increasing the number of detected $\gamma$-ray pulsars
from seven to more than sixty 
(e.g., Abdo et al.~2010a).
The {\it AGILE} has also reported the detection of about twenty 
$\gamma$-ray pulsars 
(e.g., Pellizzoni et al.~2009).
Since interpreting $\gamma$-rays should be less ambiguous
compared with reprocessed, non-thermal X-rays,
The $\gamma$-ray pulsations observed from these objects
are particularly important as a direct signature of 
basic non-thermal processes in pulsar magnetospheres,
and potentially should help to discriminate among different emission
models.

\begin{figure}
 \epsscale{1.0}
 \plotone{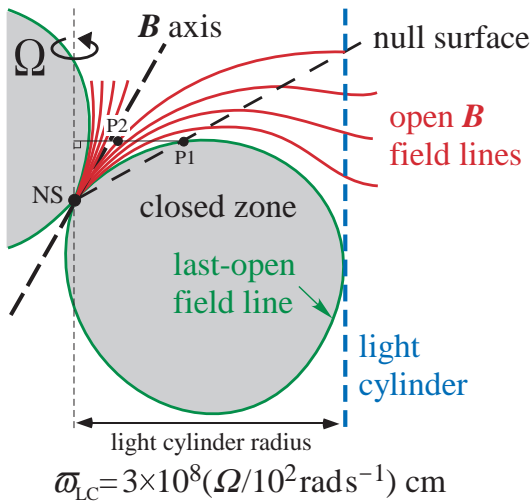}
\caption{
Side view of magnetic field lines on the meridional plane
in which both the magnetic and rotational axes reside.
\label{fig:sidev}
}
\end{figure}

In a pulsar magnetosphere (fig.~\ref{fig:sidev}), 
there is a surface called the \lq null surface', on which  
the Goldreich-Julian charge density
(Goldreich \& Julian~1969)
$\rhoGJ \equiv \mbox{\boldmath$B$}\cdot \mbox{\boldmath$\Omega$}/(2\pi c)$ 
changes sign,
where $\mbox{\boldmath$B$}$ denotes the local magnetic field, 
$\mbox{\boldmath$\Omega$}$ the NS angular-velocity vector, and
$c$ the speed of light.
There is another characteristic surface called the \lq light cylinder' 
beyond which the co-rotational velocity exceeds $c$. 
The distance of the light cylinder from the rotation axis
is called the \lq light-cylinder radius',
$\rlc \equiv c/\Omega$, 
where $\Omega \equiv \vert \mbox{\boldmath$\Omega$} \vert$ denotes
the NS rotational angular frequency.
On each magnetic azimuthal angle (measured around the magnetic axis),
there is a magnetic field line that crosses the light cylinder
tangentially; 
they are called the \lq the last-open field lines'.

In the lower colatitudes than the last-open field lines,
field lines closes inside the light cylinder;
thus, a high plasma density is expected by the trapping
of relativistic charges due to magnetic mirrors,
preventing the occurrence of a strong 
magnetic-field-aligned electric field, $\Ell$, in this closed zone.
In the higher colatitudes, on the other hand,
field lines open to a distant region beyond the light cylinder,
allowing a pulsar wind to flow along them,
thereby resulting in a high vacuum
with unscreened $\Ell$ in some regions in this open-field-line zone.

On these grounds,
in all the pulsar emission models, 
particle acceleration is assumed to realize in the open zone.
In polar-cap (PC) models, emission takes place
within a few neutron star (NS) radii above a PC surface
(Arons \& Scharlemann 1979; Daugherty \& Harding 1982, 1996).
However, the observed cutoff energies,
typically between 1~GeV and 4~GeV (Abdo et al.~2010a),
along with the 25~GeV detection of pulsed signals from the Crab pulsar
(Aliu at al.~2008),
suggests no attenuation from one-photon absorption
by a strong magnetic field (Baring 2004) near the NS,
indicating the minimum altitudes of a few stellar radii
above the PC surface.
Thus, the possibility of a high-altitude emission
gathered attention.
%

Recent pulsar high-energy emission models adopt, therefore, 
higher-altitude geometries.
There are two main scenarios in this approach:
the outer-gap (OG) model 
(Cheng et al.~1986a, b; Romani~1996; Cheng et al.~2000; 
 Hirotani~2008; Takata et al.~2008; 
 Romani \& Watters~2010), 
and the pair-starved polar-cap model
(Frackowiak \& Rudak~2005; Harding et al.~2005; Venter et al.~2009).

In the present paper, we focus on the OG model,
in which pairs are created in the outer magnetosphere
mainly by photon-photon ($\gamma$-$\gamma$) pair production.
The produced pairs polarize owing to 
the magnetic-field-aligned electric field, $\Ell$.
If the rotation and magnetic axes reside in the same hemisphere,
a positive $\Ell$ is exerted to accelerate $e^+$'s (or $e^-$'s) 
outward (or inwards), increasing the real charge density outwards.
The inward-migrating, relativistic $e^-$'s radiate 
curvature $\gamma$-rays, 
some of which (nearly head-on) collide with the X-rays 
emitted from the NS surface to materialize as pairs.

%

The purpose here is to examine the detectability of the 
$\gamma$-ray photons from pulsars with OG accelerators. 
On the neutron-star period ($P$) versus period-derivative 
($\dot{P}$) plane,
we find two kinds of death lines below which the pulsar OGs
become inactive.
In \S~\ref{sec:pair},
we derive the first kind death line,
which turns out to be a modification of 
the previous death lines (Zhang et al.~2004).
We then derive the second kind, new death line
in \S~\ref{sec:threshold},
by separately considering 
the threshold of $\gamma$-$\gamma$ pair production.
We finally combine these two death lines and
discuss the detectability of pulsars 
with recent $\gamma$-ray telescopes.

\section{Death Line derived by multiplicity constraint}
\label{sec:pair}
According to the analytical consideration in classic OG model
(Cheng et al.~1986a, b; Romani~1996; Cheng et al.~2000),
and to the numerical solution of the OG
(Hirotani~2006a,b),
an OG is found to exist essentially between the null surface and the 
light cylinder (fig.~\ref{fig:sidev}).
Since most of the pairs are produced in the inner-most region of 
the OG, the (outward-moving) $e^+$'s propagate longer distance 
than the (inward-moving) $e^-$'s.
As a result, the outward $\gamma$-ray flux becomes greater than 
the inward one.
This result is particularly important when we predict the 
high-energy emission properties from an OG.
Nevertheless, to investigate the condition for an OG to be
self-sustained, 
it is crucial to consider the collisions between
the {\it inward} $\gamma$-rays, which are emitted by $e^-$'s,
and the outward-propagating X-rays,
which are emitted from the heated PC surface.
Thus, in this paper, we concentrate on the
condition for the inward-propagating $\gamma$-rays 
to materialize as pairs
in the inner magnetosphere.

For young pulsars like Crab or Vela,
$\gamma$-$\gamma$ pair production is sustained by the
strong thermal emission from the cooling NS surface, 
and the gap trans-field thickness is kept thin.
However, as the NS ages, 
the heated PC emission due to the bombardment of in-falling 
$e^-$'s dominates the cooling NS emission
and controls the gap trans-field thickness.
Therefore, when we investigate the death lines below which 
an OG accelerator ceases to emit $\gamma$-rays,
it is appropriate to consider the heated PC as the source of
X-rays for $\gamma$-$\gamma$ pair production.
For simplicity, we assume that the heated PC has a uniform
temperature $kT$ with area $\pi R_{\rm pc}^2$.
When the OG activity is marginally sustained,
pair production takes place mainly near the null surface.


\subsection{Optical depth of photon-photon pair production}
\label{sec:tau}
At an altitude $r-r_\ast \gg R_{\rm pc}$,
the photon energy flux is given by
\begin{equation}
  F= F_\ast \left(\frac{R_{\rm pc}}{r-r_\ast}\right)^2
    \cos\Theta,
  \label{eq:flux}
\end{equation}
where the flux measured at the surface becomes
\begin{equation}
  F_\ast= 1.0 \times 10^{20} \left(\frac{kT}{100\mbox{\,eV}}\right)^4
  \mbox{\,erg\,s}^{-1}\mbox{\,cm}^{-2};
  \label{eq:flux0}
\end{equation}
$r_\ast$ denotes the NS radius, 
$\Theta$ the colatitude of the point from the {\it magnetic} axis.
In this paper, we consider that the magnetic axis is moderately or 
highly inclined with respect to the rotation axis.
In this case, the PC surface is almost face-on from the gap inner 
boundary, which is located well inside the light cylinder.
Thus, we adopt $\cos\Theta \approx 1$ in what follows.

Evaluating the photon number flux at typical blackbody
photon energy, $h\nu_{\rm x} \approx 2.8 kT$, we obtain the mean-free path
for an inward-propagating $\gamma$-ray to materialize
in a collision with the surface X-ray as
\begin{eqnarray}
  \lefteqn{
  \lambda_{\rm p}
  = \left(\frac{1}{c}\frac{F}{h\nu_{\rm x}}\right)^{-1}
    \sigma_{\rm p}^{-1}
  }
  \nonumber\\
  & & \approx 9.9 \times 10^5 
                    \left(\frac{kT}{100\mbox{\,eV}}\right)^{-3}
                    \left(\frac{r-r_\ast}{R_{\rm pc}}\right)^2
                    \mbox{\,\,cm\,},
      \qquad
  \label{eq:mfp}
\end{eqnarray}
where the $\gamma$-$\gamma$ pair-production total cross section 
is evaluated by
$\sigma_{\rm p}\sim 0.2\sigma_{\rm T}
 = 1.3 \times 10^{-25} \mbox{\,cm}^2$,
because the collisions take place mostly head-on.

Inward-propagating $\gamma$-rays are typically emitted 
from the central colatitude in the gap.
Therefore, before crossing the critical field lines
on which the null surface intersects the light cylinder
(i.e., before escaping from the gap),
they propagate the distance
\begin{equation}
  l \approx \frac{r_{\rm null}}{\sqrt{2}}
            \left( \sin\theta_{\rm null}
                  -\tan\alpha\cos\theta_{\rm null}
            \right),
  \label{eq:length}
\end{equation}
where $r_{\rm null}$ denotes the distance of the null surface
on the last-open field line from the NS center,
$\alpha$ the angle between rotation and magnetic axes.
If $\alpha=75^\circ$,
$\theta_{\rm null} \approx 85^\circ$ gives
$l \approx 0.67 r_{\rm null}$, whereas 
if $\alpha=45^\circ$,
$\theta_{\rm null} \approx 74^\circ$ gives
$l \approx 0.47 r_{\rm null}$, for example.
Note that the gap is supposed to exist only below the critical field lines,
because such solutions are being obtained by three-dimensional OG simulations.
Because of this geometry, the obtained death lines 
give tighter constraints than previous ones,
which are obtained by assuming that the OG occupies the entire 
open-field-line regions between the null surface and the light cylinder
(i.e., by imposing $f<1$ as in Zhang et al.~2004).

Combining equations~(\ref{eq:mfp}) and (\ref{eq:length}),
we can compute the optical depth 
$\tau = l/\lambda_{\rm p}$
for $\gamma$-$\gamma$ pair production, 
\begin{equation}
  \tau
  \approx 1.0 \left(\frac{kT}{100\mbox{\,eV}}\right)^3
              \left(\frac{R_{\rm pc}}{r_\ast}\right)^2
              \frac{\eta}{(\eta-1)^2} r_{\ast,6}
              \frac{l}{r_{\rm null}},
  \label{eq:tau0}
\end{equation}
where $\eta \equiv r/r_\ast$, 
$r_{\ast,6} \equiv r/(10^6 \mbox{cm})$,
$P=2\pi/\Omega$ is the NS spin period; 
$r \approx r_{\rm null}$ is used.

\subsection{Acceleration electric field}
\label{sec:Ell}
For the pulsars near the death line (to be derived below),
the potential drop in the OG
becomes a good fraction of the electromotive force,
$V_\ast \approx B_\ast r_\ast^3/(2 \rlc^2)$,
exerted on the PC surface.
Thus, we can reasonably evaluate the maximally possible $\Ell$ by
\begin{equation}
  \Ell
  \approx \frac{0.5 V_\ast}{\rlc}
  \approx \frac{B_\ast r_\ast^3}{4\rlc^3}
  = \frac{\mu}{2 \rlc^3},
  \label{eq:Epara}
\end{equation}
where 
$\mu$ denotes the NS dipole moment,
and $B_\ast$ the magnetic field strength at the NS surface.
Since the potential drop occurs only in a portion of the open zone, 
we multiply the factor $0.5$ in the numerator.
By this acceleration electric field, 
the inward motion of $e^-$'s attain a force balance
between the electrostatic and curvature-radiation forces.
Thus, the typical Lorentz factors of the $e^-$'s become
\begin{equation}
  \gamma = \left(\frac{3 R_{\rm c}^2}{2e} E_\parallel \right)^{1/4},
  \label{eq:Lf}
\end{equation}
where $e$ refers to the magnitude of the charge on the electron,
$R_{\rm c}$ the curvature radius of the magnetic field line.

\subsection{Evaluation of curvature radius}
\label{sec:cases}
To evaluate $R_{\rm c}$, we we consider the following 
two representative cases.\\
{\bf Case 1:} The null surface intersects the last-open field line
in the {\it inner-most} magnetosphere; that is, 
$r_{\rm null} < 0.05 \rlc$,
In this case, the curvature radius can be approximated as
\begin{equation}
  R_{\rm c} \approx \frac43 \sqrt{\frac{r}{\rlc}}\rlc
  \label{eq:Rc1}
\end{equation}
on the last-open field line.
If $\alpha > 70^\circ$, 
this case becomes applicable along the magnetic field lines
that crosses the NS surface with magnetic azimuthal angle 
in the range $-60^\circ < \varphi_\ast < 60^\circ$,
where $\varphi_\ast$ increases counter-clockwise and
$\varphi_\ast=\pi$ points the rotation axis from the magnetic axis.
To derive the OG death line, 
it is sufficient to consider only this range,
$\vert\varphi_\ast\vert < 60^\circ$,
because the OG is active only in 
$\vert\varphi_\ast\vert < 60^\circ$.
\\
{\bf Case 2:} The null surface intersects the last-open field line
in the {\it outer} magnetosphere; that is, 
$r_{\rm null} > 0.07 \rlc$.
In this case, we have
\begin{equation}
  R_{\rm c} \approx 0.5 \rlc.
  \label{eq:Rc2}
\end{equation}
If $\alpha<62^\circ$,
this case becomes applicable
in $\vert\varphi_\ast\vert < 60^\circ$.

\subsection{Typical gamma-ray energies}
\label{sec:energy}
Substituting equation~(\ref{eq:Epara}) into (\ref{eq:Lf}),
we obtain
\begin{equation}
  \gamma
  \approx 3.3 \times 10^6 \, \eta^{1/4} 
                            \, P^{-1/2} 
                            \, \mu_{30}{}^{1/4}
                            \, r_{\ast,6}{}^{1/4},
  \label{eq:Lf1}
\end{equation}
for case~1, and 
\begin{equation}
  \gamma
  \approx 1.7 \times 10^7 \, P^{-1/4} 
                            \, \mu_{30}{}^{1/4},
  \label{eq:Lf2}
\end{equation}
for case~2, where $\mu_{30}= \mu/(10^{30}\,\mbox{G cm}^3)$.

The flux density of the curvature spectrum
peaks at the frequency $0.29 \nu_{\rm c}$,
where the critical frequency is defined by
\begin{equation}
  \nu_{\rm c}\equiv \frac{3 c \gamma^3}{4\pi R_{\rm c}}.
  \label{eq:nu_c}
\end{equation}
However, for the pulsars near the death line,
$\gamma$-$\gamma$ pair production is maintained by the 
collisions between the $\gamma$-ray photons 
with energy $h\nu_\gamma \gg 0.29 h\nu_{\rm c}$
and the thermal photons near the blackbody peak 
$h\nu_{\rm s} \sim 2.82 kT$, or
between the $\gamma$-rays with $h\nu_\gamma \sim 0.29 h\nu_{\rm c}$
and the thermal photons in the Wien regime $h\nu_{\rm s} \gg 2.82 kT$,
or both (i.e., $h\nu_\gamma > 0.29 h\nu_{\rm c}$ and 
$h\nu_{\rm s} > 2.82 kT$).
To take account of such effects, we can practically assume
that pair production takes place
between the $\gamma$-rays with $h\nu_\gamma \sim h\nu_{\rm c}$
and the surface X-rays with $h\nu_{\rm s} \sim 3 kT$;
this treatment incurs relatively small errors from
the exact, complicated numerical computations.

Substituting equations~(\ref{eq:Lf1}) or (\ref{eq:Lf2})
into (\ref{eq:nu_c}),
we obtain the typical $\gamma$-rays energy,
\begin{equation}
  h\nu_{\rm c}
  = 11.8 \,\eta^{1/4}
         \,P^{-2}
         \,\mu_{30}{}^{3/4}
         \,r_{\ast,6}{}^{1/4}
    \mbox{MeV},
  \label{eq:hnu_c1}
\end{equation}
\begin{equation}
  h\nu_{\rm c}
  = 60   \,P^{-7/4}
         \,\mu_{30}{}^{3/4}
    \mbox{MeV}
  \label{eq:hnu_c2}
\end{equation}
for cases~1 and 2, respectively.

\subsection{Number of gamma-rays emitted by each electron}
\label{sec:number}
Let us now evaluate number of photons emitted by a single
inward-accelerated $e^-$ near the null surface,
the inner-most region of the gap.
If $R_{\rm c}$ were constant in the magnetosphere, 
the photons emitted between
the altitudes $r_{\rm null}$ and ($r_{\rm null}+l$) illuminate 
the null surface, 
where $l$ is given by equation~(\ref{eq:length}).
Therefore, we could evaluate the number of $\gamma$-rays
emitted by a single $e^-$ by
\begin{equation}
  N_\gamma 
  \approx \frac{e E_\parallel l}{h\nu_{\rm c}}.
  \label{eq:Ngamma0}
\end{equation}                            
It follows that a single $e^-$ cascade into 
$N_\gamma \tau$ secondary pairs within the gap. 

Substituting equations~(\ref{eq:hnu_c1}) or (\ref{eq:hnu_c2})
into (\ref{eq:Ngamma0}), we obtain
\begin{equation}
  N_\gamma \approx 78\, 
              \,\eta^{3/4}
              \,P^{-1}
              \,\mu_{30}{}^{1/4}
              \,r_{\ast,6}{}^{3/4}
              \,\frac{l}{0.67 r_{\rm null}},
  \label{eq:Ngamma1}
\end{equation}
\begin{equation}
  N_\gamma \approx 10.8 \, 
              \,\eta
              \,P^{-5/4}
              \,\mu_{30}{}^{1/4}
              \,r_{\ast,6}
              \,\frac{l}{0.47 r_{\rm null}} \qquad
  \label{eq:Ngamma2}
\end{equation}
for cases 1 and 2, respectively.

\subsection{Number of secondary pairs cascaded from a primary electron}
\label{sec:secondary}
Combining equations~(\ref{eq:tau0}) and (\ref{eq:Ngamma1}),
we obtain the number of pairs that a single $e^-$ cascades
within the gap, 
\begin{eqnarray}
  N_\gamma \tau 
  &\approx& 52  \frac{\eta^{7/4}}{(\eta-1)^2}
                \left(\frac{R_{\rm pc}}{r_\ast}\right)^2
                \left(\frac{kT}{100\mbox{\,eV}}\right)^3
  \nonumber\\
  & & \times  \, P^{-1}
              \, \mu_{30}{}^{1/4}
              \, r_{\ast,6}{}^{7/4}
              \left(\frac{l}{0.67 r_{\rm null}}\right)^2
  \quad
  \label{eq:pairs1}
\end{eqnarray}
for case 1.
In the same manner, 
combining equations~(\ref{eq:tau0}) and (\ref{eq:Ngamma2}),
we obtain
\begin{eqnarray}
  N_\gamma \tau 
  &\approx& 5.0 \frac{\eta^2}{(\eta-1)^2}
                \left(\frac{R_{\rm pc}}{r_\ast}\right)^2
                \left(\frac{kT}{100\mbox{\,eV}}\right)^3
  \nonumber\\
  & & \times  \, P^{-5/4}
              \, \mu_{30}{}^{1/4}
              \, r_{\ast,6}{}^2
              \left(\frac{l}{0.47 r_{\rm null}}\right)^2
  \quad
  \label{eq:pairs2}
\end{eqnarray}
for case 2.
We have to express $(R_{\rm pc}/r_\ast)^2$ 
so that the emitted energy in the form of 
the blackbody radiation from the heated PC surface
may be consistent with the energy deposited by
the bombardment of gap-accelerated $e^-$'s.

\subsection{Effective area of heated polar-cap}
\label{sec:heatedPC}
The number flux of 
$e^-$'s falling onto the polar cap surface
(with area $A_{\rm gap}$) can be evaluated by
\begin{eqnarray}
  \dot{N}_{\rm e}
  &=& \kappa \frac{\Omega B_\ast}{2\pi e} A_{\rm gap}
  \nonumber\\
  &\approx& 
      5.5 \times 10^{29} \kappa
      P^{-2} \mu_{30} \frac{A_{\rm gap}}{0.2 A_{\rm open}},
  \label{eq:dot_Ne}
\end{eqnarray}
where $A_{\rm open} \sim \pi r_\ast^3/\rlc$ denotes the entire PC area.
Since typically $20$~\% of the open flux tubes
thread the gap, $A_{\rm gap} \sim 0.2 A_{\rm open}$
is a good compromise.
See figure~\ref{fig:PC} for an example of active field lines
on the PC surface, which indicates that about $20\%$ of the 
open field line fluxes are active. 

From numerical analysis, $\kappa \sim 0.3$ is appropriate for 
$\alpha \sim 75^\circ$, while $\kappa \sim 0.7$ for 
$\alpha \sim 45^\circ$.
In Newtonian approximation, $\kappa$ typically becomes $\cos\alpha$,
because we obtain $B_z/B=\cos\alpha$ at the PC surface,
where $B_z$ refers to the magnetic field component projected
along the rotation axis.
Thus, $\kappa$ decreases with increasing $\alpha$.
However, as the NS rapidly rotates, the space-time
dragging effect (Muslimov \& Tsygan 1992) leads to a
smaller $\kappa$ than $\cos\alpha$.  On the other hand,
in a three-dimensional pulsar magnetosphere, inward 
$\gamma$-rays preferentially propagates towards the
leading side due to the aberration of the photon propagation 
direction; thus, created current becomes super Goldreich-Julian
in the leading side to exert a space-charge-limited flow of ions 
from the PC surface in an OG (H06a). Thus, $\kappa$ can 
exceed the value $\rho_{\rm GJ}/(\Omega B / 2 \pi e)$ at the PC surface.
For a highly inclined case, the fraction of the
field lines having super-Goldreich-Julian current density
increases because of the decreased $\rho_{\rm GJ}$ at the
PC surface. Thus, $\kappa \sim 0.3$ is more appropriate
than $\cos(75^\circ)=0.26$ for $\alpha=75^\circ$.
However, for $\alpha=45^\circ$, most field lines have
sub-Goldreich-Julian current; thus, we obtain
$\kappa \approx \cos(45^\circ)= 0.70$.
As an intermediate case, 
we present the solution of $\kappa$ in a three-dimensional OG 
with $\alpha=60^\circ$ in the left panel of figure~\ref{fig:PC}, 
which shows that the evaluation of 
$\kappa \approx \cos(60^\circ)=0.5$ is appropriate on average.

Each $e^-$ brings the energy
$\gamma_{\rm h} m_{\rm e} c^2$ on the NS surface,
where $\gamma_{\rm h}$ denotes the Lorentz factor at
the altitude $r=r_{\rm h} \equiv \sqrt{2 r_\ast \rlc}$ 
below which all the emitted photons hit the NS. 
For case~1, most of the $e^-$'s kinetic energy at the gap inner boundary
will be eventually turned into the NS surface emission, 
because $r_{\rm h}/\rlc \approx 0.028 \sqrt{r_{\ast,6}/P}$
shows that $r_{\rm null} < r_{\rm h}$
if $P < 0.3$~s.
For case~2, on the contrary,
only a portion of $\gamma m_{\rm e} c^2$ at the null surface
will be used in the heating of the NS surface,
because $r_{\rm null} > r_{\rm h}$.

For case~1, we can put $\gamma_{\rm h}=\gamma$, 
where the right-hand side is evaluated by equation~(\ref{eq:Lf1}).
Thus, combining equations~(\ref{eq:Lf1}) and (\ref{eq:dot_Ne}),
we obtain
\begin{eqnarray}
  \lefteqn{
  \gamma_{\rm h} m_{\rm e} c^2 \dot{N}_{\rm e}
  \approx 4.5 \times 10^{29} \eta^{1/4} P^{-5/2} \mu_{30}^{5/4}
  }
  \nonumber\\
  &\times&  r_{\ast,6}^{1/4} \frac{\kappa}{0.3}
            \frac{A_{\rm gap}}{0.2 A_{\rm open}} \mbox{\, erg s}^{-1}.
\end{eqnarray}

For case~2, on the other hand, 
electrons lose energy by curvature radiation
from $r=r_{\rm null}$ down to $r=r_{\rm h}$
before illuminating the NS.
Integrating the equation of Lorentz factor evolution,
\begin{equation}
  \frac{d\gamma}{dt} m_{\rm e} c^2
  = \frac{2 e^2}{3 R_{\rm c}^2} \gamma^4
  \label{eq:Lf_evol}
\end{equation}
from $r=r_{\rm null}$ to $r=r_{\rm h}$, 
we obtain
\begin{equation}
  \gamma_{\rm h} 
  \approx 2.5 \times 10^7 P^{1/3} 
          \left[\ln\left(\frac{r_{\rm null}}{r_{\rm h}}\right)
          \right]^{-1/3}
  \label{eq:Lf_evol2}
\end{equation}
where the weak dependence on $r_{\rm null}/r_{\rm h}$,
which is of the order of unity, may be neglected.
Combining equations~(\ref{eq:Lf_evol2}) and (\ref{eq:dot_Ne}),
we obtain
\begin{eqnarray}
  \gamma_{\rm h} m_{\rm e} c^2 \dot{N}_{\rm e}
  &\approx& 7.8 \times 10^{30} P^{-5/3} \mu_{30}
  \nonumber\\
  & & \times  \frac{\kappa}{0.7}
              \frac{A_{\rm gap}}{0.2 A_{\rm open}} \mbox{\, erg s}^{-1}.
\end{eqnarray}

Equating the heating rate
$\gamma_{\rm h} m_{\rm e} c^2 \dot{N}_{\rm e}$
derived just above
with the emission rate
$\pi R_{\rm pc}{}^2 \sigma T^4$,
where $\sigma$ denotes the Stefan-Boltzmann constant,
we can evaluate the effective area of the heated PC region,
$\pi R_{\rm pc}^2$, and obtain
\begin{eqnarray}
  \lefteqn{
    \left(\frac{R_{\rm pc}}{r_\ast}\right)^2
    = 1.4 \times 10^{-3}
        P^{-5/2} \mu_{30}{}^{5/4}
          }
  \nonumber\\
  & & \times 
      \left(\frac{kT}{100\mbox{\,eV}}\right)^{-4} 
      \eta^{1/4} 
      r_{\ast,6}^{-7/4} 
      \frac{\kappa}{0.3}
      \frac{A_{\rm gap}}{0.2 A_{\rm open}}
  \quad
  \label{eq:Rfac1}
\end{eqnarray}
for case 1, and 
\begin{eqnarray}
  \lefteqn{
    \left(\frac{R_{\rm pc}}{r_\ast}\right)^2
      = 2.4 \times 10^{-2}
        P^{-5/3} \mu_{30}
          }
  \nonumber\\
  & & \times 
      \left(\frac{kT}{100\mbox{\,eV}}\right)^{-4}
      r_{\ast,6}^{-2} 
      \frac{\kappa}{0.7}
      \frac{A_{\rm gap}}{0.2 A_{\rm open}}
  \quad
  \label{eq:Rfac2}
\end{eqnarray}
for case 2.
Note that $A_{\rm gap}/A_{\rm open}$ merely parametrizes the
fraction of active field lines to the entire open field lines.

\subsection{Multiplicity constraint}
\label{sec:death line}
For case~1, substituting equation~(\ref{eq:Rfac1}) into
(\ref{eq:pairs1}), we obtain
\begin{eqnarray}
  \lefteqn{
    N_\gamma \tau 
    \approx 6.3 \times 10^{-2}
                \left(\frac{\eta}{\eta-1}\right)^2
                \left(\frac{kT}{100\mbox{\,eV}}\right)^{-1}
          }
  \nonumber\\
  &\times&   P^{-7/2}
              \mu_{30}^{3/2}
              \frac{\kappa}{0.3}
              \frac{A_{\rm gap}}{0.2 A_{\rm open}}
              \left(\frac{l}{0.67 r_{\rm null}}\right)^2.
  \quad
  \label{eq:pairs1b}
\end{eqnarray}

For case~2, substituting equation~(\ref{eq:Rfac2}) into
(\ref{eq:pairs2}), we obtain
\begin{eqnarray}
  \lefteqn{
    N_\gamma \tau 
    \approx 1.2 \times 10^{-1}
                \left(\frac{\eta}{\eta-1}\right)^2
                \left(\frac{kT}{100\mbox{\,eV}}\right)^{-1}
          }
  \nonumber\\
  &\times&   P^{-35/12}
              \mu_{30}^{5/4}
              \frac{\kappa}{0.7}
              \frac{A_{\rm gap}}{0.2 A_{\rm open}}
              \left(\frac{l}{0.47 r_{\rm null}}\right)^2.
  \quad
  \label{eq:pairs2b}
\end{eqnarray}

In this paper, we evaluate the NS magnetic moment
assuming the vacuum dipole radiation formula,
$\mu^2= 3 I c^3 P \dot{P} /(8\pi^2)$.
Then, equations (\ref{eq:pairs1b}) and (\ref{eq:pairs2b})
give
\begin{eqnarray}
  \lefteqn{
    N_\gamma \tau 
    \approx 7.4 \times 10^{-2}
                \left(\frac{\eta}{\eta-1}\right)^2
                \left(\frac{kT}{100\mbox{\,eV}}\right)^{-1}
                I_{45}{}^{3/4}
          }
  \nonumber\\
  &\times&
              P^{-11/4}
              \dot{P}_{-15}{}^{3/4}
              \frac{\kappa}{0.3}
              \frac{A_{\rm gap}}{0.2 A_{\rm open}}
              \left(\frac{l}{0.67 r_{\rm null}}\right)^2,
  \qquad
  \label{eq:pairs1c}
\end{eqnarray}
\begin{eqnarray}
  \lefteqn{
    N_\gamma \tau 
    \approx 1.2 \times 10^{-1}
                \left(\frac{\eta}{\eta-1}\right)^2
                \left(\frac{kT}{100\mbox{\,eV}}\right)^{-1}
                I_{45}{}^{5/8}
          }
  \nonumber\\
  &\times&
              P^{-55/24}
              \dot{P}_{-15}{}^{5/8}
              \frac{\kappa}{0.7}
              \frac{A_{\rm gap}}{0.2 A_{\rm open}}
              \left(\frac{l}{0.47 r_{\rm null}}\right)^2
  \qquad
  \label{eq:pairs2c}
\end{eqnarray}
for cases~1 and 2, respectively.

Setting $N_\gamma \tau > 1$, we obtain the death line
on the $P$ versus $\dot{P}$ plane,
which describes the condition for an OG to emit $\gamma$-rays efficiently.
Thus, the minimum spin-down rate for a given $P$ is obtained as
\begin{equation}
  \lg\dot{P} = -13.49 + 3.67 \lg P,
  \label{eq:death line_1a}
\end{equation}
\begin{equation}
  \lg\dot{P} = -13.54 + 3.67 \lg P
  \label{eq:death line_2a}
\end{equation}
for cases~1 and 2, respectively.

Since we consider that the gap will not extend in the entire
open field line region, 
the obtained death lines~(\ref{eq:death line_1a}) and 
(\ref{eq:death line_2a}) gives a tighter constraint than
previous works.
For example, Zhang et al.~(2004) gives
\begin{equation}
  \lg\dot{P} = -14.60 + 3.33 \lg P
  \label{eq:death line_3a}
\end{equation}
for a uniform distribution of $\alpha$'s, and
\begin{equation}
  \lg\dot{P} = -14.20 + 3.33 \lg P
  \label{eq:death line_3b}
\end{equation}
for a cosine distribution of $\alpha$.
Death Lines~(\ref{eq:death line_3a}) and (\ref{eq:death line_3b})
are depicted by the thick and thin dash-dotted lines
in figure~\ref{fig:deathline}.

\section{Death Line derived by pair-production threshold}
\label{sec:threshold}
We now turn to the second and important constraint obtained by
pair-production threshold.
To derive this, we must evaluate the maximally possible
energies of both the surface X-rays and the curvature $\gamma$-rays.

First, let us examine the maximum surface temperature,
$kT_{\rm max}$, 
which is probably realized in a limited area of the heated PC,
near the footpoints of active magnetic field lines.
That is, we release the assumption of uniform $kT$ on the heated
PC surface, which was adopted in \S~\ref{sec:pair}.
Electrons will fall onto the NS surface with the typical 
Lorentz factor given by equation~(\ref{eq:Lf_evol2}),
where $r_{\rm h}$ may be replaced with $r_\ast$.
However, the logarithmic factor little changes 
the right-hand side; thus, we evaluate the Lorentz factor
of in-falling $e^-$'s by
$\gamma_7= 2.5 P^{1/3}$,
where $\gamma_7= \gamma/10^7$.
Equating $\sigma T_{\rm max}{}^4$ with 
$ \gamma m_{\rm e} c^2 \dot{N}_{\rm e} / A_{\rm gap}
 =\gamma m_{\rm e} c^2 \kappa \Omega B_\ast / (2\pi e)$,
we obtain
\begin{equation}
  kT_{\rm max}
  = 426 \mbox{\,eV} \kappa^{1/4} \gamma_7^{1/4}
    P^{-1/4} \mu_{30}^{1/4} r_{\ast,6}{}^{-3/4}.
  \label{eq:kTmax}
\end{equation}
Therefore, the typical X-ray energy can be estimated by
\begin{eqnarray}
  h \nu_X 
  &\approx& 
  3 k T_{\rm max}
  \nonumber\\
  &\approx& 
  0.0023 m_{\rm e} c^2 
                  \left(\frac{\kappa}{0.3}\right)^{1/4}
                  P^{-1/6} \mu_{30}{}^{1/4}
                  r_{\ast,6}{}^{-3/4}
  \qquad
\end{eqnarray}
for both case~1 and case~2, 
apart from the difference of $\kappa$.

Second, we examine the maximum energy of curvature photons.
For case~1, we use equation~(\ref{eq:hnu_c1}) with
$\eta = 2.4 \times 10^3 P (r/0.5\rlc)$
to obtain
\begin{equation}
  h\nu_\gamma
  = 1.6 \times 10^2 m_{\rm e} c^2 
    P^{-7/4} \mu_{30}^{3/4} r_{\ast,6}{}^{1/4}.
  \label{eq:Eg}
\end{equation}
For case~2, we compute $h\nu_\gamma$ by equation~(\ref{eq:hnu_c2}).

Third, we impose the pair-production threshold condition,
$h \nu_X \times h\nu_\gamma > (m_{\rm e} c^2)^2$
to obtain
\begin{equation}
  \lg\dot{P} = -14.15 + 2.83 \lg P,
  \label{eq:death line_1b}
\end{equation}
\begin{equation}
  \lg\dot{P} = -14.06 + 2.83 \lg P
  \label{eq:death line_2b}
\end{equation}
for cases~1 and 2, respectively. 

\begin{figure}
 \epsscale{1.15}
 \plotone{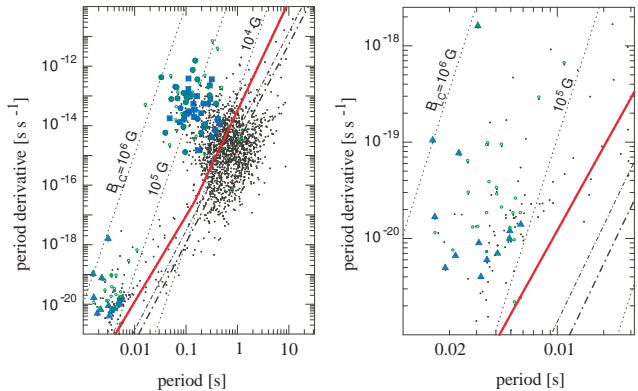}
\caption{
Outer-gap death line on the period versus period-derivative plane.
The (red) solid line denotes the death line 
obtained in the present work.
For comparison, previous outer-gap death lines are
depicted as the thick and thin dash-dotted lines
(Zhang et al.~2004).
The thin dotted lines expresses the parameter region that gives
the labeled magnetic-field strength, $B_{\rm LC}$, 
at the light cylinder.
The (blue) large points denote $\gamma$-ray detected pulsars, whereas 
the (green) small open circles denote X-ray detected pulsars
(see text for details and references).
{\it Left}: entire parameter space.
{\it Right}: close up of millisecond parameter range.
\label{fig:deathline}
}
\end{figure}

On these grounds, the actual death line is obtained by taking the greater
$\dot{P}$ in each case.
For case~1, we obtain
\begin{equation}
  \lg\dot{P} = \max(-13.49 + 3.67 \lg P,-14.15 + 2.83 \lg P),
  \label{eq:death line_1}
\end{equation}
and for case~2, we obtain
\begin{equation}
  \lg\dot{P} = \max(-13.54 + 3.67 \lg P,-14.06 + 2.83 \lg P).
  \label{eq:death line_2}
\end{equation}
It follows that the death line little depends on $\alpha$.
In figure~\ref{fig:deathline}, 
the death line~(\ref{eq:death line_1}) 
is plotted as the (red) thick solid line.
Case~2~(eq.~[\ref{eq:death line_2}]) is not plotted,
because it almost overlaps on case~1~(eq.~[\ref{eq:death line_1}]).
The (blue) squares denote the $\gamma$-ray detected pulsars,
the (blue) triangles the millisecond $\gamma$-ray pulsars,
and the (blue) circles other radio-loud $\gamma$-ray pulsars
detected with Fermi and/or AGILE
(Aliu et al.~2008; Halpern et al.~2008; 
 Pellizzoni et al.~2009; 
 Abdo et al.~2010a; Abdo et al.~2010b; Abdo et al.~2010c;
 Pilia et al.~2010; Saz Parkinson et al.~2010; 
 Pilia \& Pellizzoni~2011; Saz Parkinson et al.~2011;
 Guillemot et al.~2011; Keith et al.~2011). 
The (green) small open circles denotes the X-ray detected pulsars
(see Grindlay and Bogdanov 2009,
 Becker \& Tr\"umper 1999,
 Becker 2009, for reviews;
 see also Grindlay et al. 2002, Bogdanov et al. 2006,
 Bogdanov et al. 2010, Bogdanov et al. 2011, 
 Saito et al. 1997, Becker et al. 2003, Bassa et al. 2004,
 D'Amico et al. 2002, Elsner et al. 2008, and Heinke et al. 2006
 for the millisecond pulsars in globular clusters;
 Kuiper et al. 2000, Mineo et al. 2000, Kuiper et al. 2002
 for J0218+4232;
 Zavlin \& Pavlov 1998, Zavlin et al. 2002
 for J0437-4715;
 Bogdanov \& Grindlay 2009 
 for J0030+0451;
 Zavlin 2006 
 for J0437-4715, J2124-3358, J1024-0719, J0034-0534
).
The small black dots denote radio pulsars that do not show
either $\gamma$-ray or X-ray pulsations
(ATNF pulsar catalog
 \footnote{http://www.atnf.csiro.au/research/pulsar/psrcat/}).

Below the death line, 
or equivalently in the so-called \lq death valley',
the OG activity ceases.
It follows that all the $\gamma$-ray pulsars detected so far
and 84 (of 89) X-ray pulsars are located above the death line.
The remaining five X-ray pulsars in the death valley,
all of which are old nearby pulsars,
will be discussed in \S~\ref{sec:disc_old}.

It also follows from figure~\ref{fig:deathline} 
that the new constraint obtained 
by the pair-production threshold (\S~\ref{sec:threshold})
becomes important when we consider
fast rotators ($P<0.1$~s).
To see this, the close-up figure of millisecond parameter range
is presented as the right panel.
It shows that all the X-ray and $\gamma$-ray millisecond pulsars (MSPs) 
are above the death line, 
including the marginal two cases
(J1911-6000C in NGC 6752 and J0024-7204C in 47-Tuc).
If the X-ray emission of such a marginal case
is turned out to be thermal, 
it is probably due to the heated polar cap emission associated with
OG activities.
On the other hand, if they are non-thermal
(as in the case of old, nearby, non-recycled pulsars),
it may be due to a magnetospheric emission from the extended polar gap
(see \S~\ref{sec:disc_old}).

\section{Discussion}
\label{sec:discussion}
It is found that the death line has a weak dependence on $\alpha$.
This is because negative feedback effects work 
(Hirotani~2006) in an OG and its electrodynamic structure
little changes when we moderately change such parameters as
$\alpha$, $P$, and $\mu$.
However, the gap electrodynamic structure is mildly subject to change
when we adopt different $kT$ 
(eqs.~[\ref{eq:pairs1c}] \& [\ref{eq:pairs2c}]),
because it directly affects the efficiency of 
$\gamma$-$\gamma$ pair production.

In their death line calculation,
Zhang et al.~(2004) did consider 
the $\gamma$-$\gamma$ pair-production threshold 
and wrote down the fractional gap thickness, $f$,
as a function of $\alpha$, $P$, $\dot{P}$, and $r$.
In this paper, instead of using $f$,
we considered two conditions
($N_\gamma \tau > 1$ and pair-production threshold)
to find that the latter condition
gives a tighter constraint for MSPs.
The difference becomes particularly important when we investigate
potential MSPs in globular clusters,
which provide an efficient way to observe less-luminous MSPs 
with long exposures.

\subsection{Death line of millisecond pulsars}
\label{sec:disc_MSP}
For slowly rotating pulsars ($P>0.1$~s),
the OG is active if the
magnetic field strength at the light cylinder, $B_{\rm LC}$,
is greater than $10^4$~G.
However, for MSPs ($P<0.01$~s),
the OG has enough luminosity only when $B_{\rm LC}>10^5$~G.
The reasons are twofold:
lower PC temperature and soft curvature spectrum.
For the PC temperature, equation~(\ref{eq:kTmax}) gives
$kT_{\rm max} \propto B_{\rm LC}^{5/16} P^{11/16}$,
where $B_{\rm LC} \propto \dot{P}^{1/2} P^{-5/2}$
and $\mu \propto P^{1/2} \dot{P}^{1/2}$ are used.
Therefore, for the same $B_{\rm LC}$,
the heated PC temperature decreases
with decreasing $P$.
For the curvature photon energy,
equations~(\ref{eq:Epara}) and (\ref{eq:Lf}) give
$h \nu_{\rm c} \propto B_{\rm LC}^{3/4} P^{5/4}$.
Therefore, 
the maximally possible curvature photon energy, 
$h \nu_{\rm c}$, also decreases with decreasing $P$.

In spite of the facts above, we should notice here that 
the gap meridional thickness is self-regulated 
so that the pair multiplicity within the gap may be kept unity
on average.
Therefore, the curvature photon energy 
(or equivalently, the exponential-cutoff energy) 
little changes as the pulsar ages.
This argument is valid until the pulsar approaches
the death line, and will be discussed in a separate paper.

\subsection{Old nearby pulsars in the outer-gap death valley}
\label{sec:disc_old}
In figure~\ref{fig:deathline}, we plot the X-ray detected pulsars
by small (green) open circles.
It follows that several pulsars with longer periods
($P > 0.5$~s) are located in the
\lq death valley', which indicates that their OGs have already
finished activity.
They are B0823+26, B1133+16, J0108-1431, B0943+10, B0628-28,
all of which are old non-recycled pulsars with characteristic 
ages around $10^7$ years
(Becker et al. 2004; Becker et al. 2005; Becker et al. 2006).
Although the photon statistics are limited, 
B0823+26 and B0628-28 are found to show predominantly non-thermal spectra.

We consider that their non-thermal X-rays are emitted from
the PC accelerator (or polar gap) extended into the higher altitudes
along the magnetic field lines curving {\it toward} the
rotation axis 
(Fawley et al. 1977; Scharlemann et al. 1978;
 Arons \& Scharlemann 1979; Arons 1983).
For a moderately inclined rotator, 
such \lq toward curvature' filed lines exist on the opposite side 
of the OG on the PC surface
(e.g., in the magnetic azimuthal angle range 
between $105^\circ$ and $220^\circ$ in fig.~\ref{fig:PC} 
for $\alpha=60^\circ$,
measured counter-clockwise from the $+x$ direction).
In another word, after the OG in the first and fourth quadrant
ceases activity, the extended polar gap in the second and the third
quadrant remains active to emit non-thermal X-rays or soft $\gamma$-rays
with small luminosity.
Since it is reasonable to assume a free emission of $e^-$'s
from the NS surface by $\Ell$ (Arons 1981),
there will be no \lq death lines' for this kind of extended polar gap.
The space-charge-limited flow of $e^-$'s will be continuously 
accelerated outwards by a week $\Ell$ to emit 
curvature photons from X-ray to soft $\gamma$-ray energies, 
depending on the potential drop in this gap.
The photons will be emitted into the direction
$\theta<\alpha$ and $\pi-\theta<\alpha$,
where $\theta$ denotes the photon propagation angle
with respect to the rotation axis.
For instance, the total emission solid angle becomes
$2\pi$ if $\alpha=60^\circ$.
This corresponds to a modification of the original
pair-starved polar-cap (PSPC) model
attempted to apply to the \lq away curvature' field lines
(Muslimov \& Harding 2004a,b; Venter et al. 2009).
This topic, a modified version of the PSPC model, 
will be discussed in a separate paper.

\subsection{Heated polar cap of millisecond pulsars}
\label{sec:disc_Xray}
\begin{figure}
 \epsscale{1.20}
 \plotone{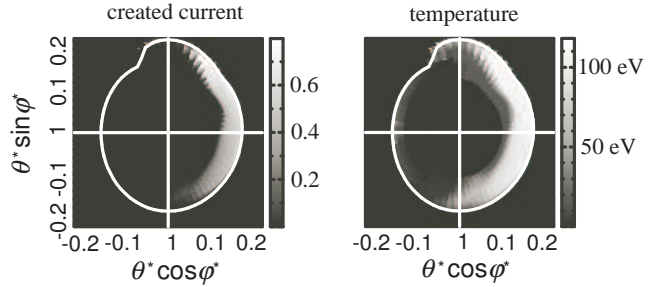} 
\caption{
Solution of a self-consistent three-dimensional outer-gap
for an example millisecond pulsar
with $P=10$~ms, $\dot{P}=10^{-19} \,\mbox{s\,s}^{-1}$,
and $\alpha=60^\circ$.
Both the abscissa and the ordinate are in radian unit.
The white curve denotes the foot points of the last-open
magnetic field lines on the polar-cap (PC) surface.
The magnetic field is assumed to be dipolar centered
at the NS center. 
{\it Left:} 
Distribution of $\kappa$ (eq.~\ref{eq:dot_Ne}), 
or equivalently, the created current density normalized by
the typical Goldreich-Julian value,
$\Omega B / (2 \pi)$ on the polar-cap surface.
{\it Right:}
Distribution of the maximum attainable temperature [eV]
owing to the bombardment of the gap-accelerated electrons
on the PC surface.
Heat conduction in the trans-field direction is not considered.
\label{fig:PC}
}
\end{figure}

Let us finally consider the relationship with X-ray observations.
To this aim, we present the OG solution obtained for a typical
millisecond pulsar parameter,
$P=10$~ms, $\dot{P}=10^{-19}\,\mbox{s\,s}^{-1}$,
by applying the numerical technique described in
Hirotani (2011).
Since the solution close to the death line cannot be
easily obtained by numerical analysis, 
we consider this set of ($P$, $\dot{P}$),
which is located relatively away from the death line 
(fig.~\ref{fig:deathline}).
To consider a moderate case, we assume $\alpha=60^\circ$.
We present the solved $\kappa$,
the created current density normalized by the 
typical Goldreich-Julian (GJ) value, $\Omega B / 2\pi$
in the left panel of figure~\ref{fig:PC}.
It follows that the created current becomes
super-GJ, $\kappa > \cos\alpha=0.5$,
in the leading side (in the first quadrant),
whereas sub-GJ in the trailing side (in the fourth quadrant).
This is because the inward $\gamma$-rays propagate
towards the rotational direction due to the aberration of the
photon propagation direction to preferentially materialize
as pairs in the leading side.
The created current concentrate near the last-open field line
in the leading side, 
because the magnetic flux surface 
(i.e., a constant $A_\varphi$ surface)
has a negative extrinsic curvature
(like the saddle or the inner surface of a daughnut)
in the lower altitudes where pair production mainly takes place.
The inward $\gamma$-rays preferentially propagate
towards the lower magnetic colatitudes 
(i.e., towards the magnetic equator)
to materialize as pairs near the last-open field line.
This forms a striking contrast with
two-dimensional calculation,
which predicts that the created current peaks in the
middle or higher magnetic colatitudes 
(i.e., away from the last-open field lines)
in the gap (fig.~1 of Takata et al. 2008).

Using $\kappa$ and the Lorentz factors of the in-falling $e^-$'s
at the PC surface, we can compute the maximum attainable
surface temperature using equation~(\ref{eq:kTmax}).
The result is presented in the right panel of figure~\ref{fig:PC}.
The integrated PC X-ray luminosity becomes 
$3.4 \times 10^{30} \,\mbox{erg\,s}^{-1}$,
whereas the magnetospheric $\gamma$-ray luminosity becomes
$4.4 \times 10^{32} \,\mbox{erg\,s}^{-1}$.
We should notice here that this figure does not represent
the actual temperature distribution on the PC surface of
a millisecond pulsar by any means.
For a weak surface magnetic field ($B<10^9$~G) of 
millisecond pulsars,
the cyclotron energy is comparable or less than the 
Coulomb energy.
Thus, we can expect more or less isotropic heat conduction
on the surface of a millisecond pulsar.
The high heat conduction of subphotospheric layers,
where the in-falling plasmas' energy is released,
allows the heat to propagate across the NS surface,
resulting in a greater hot region than given 
in figure~\ref{fig:PC} (right) 
with decreasing temperature toward the rim of the heated region.
Since there is no reliable calculation of
the heat conduction on a NS surface,
we simply present the maximum attainable temperature in this paper.

From the X-ray observations of several nearby field
millisecond pulsars
(see Pavlov \& Zavlin 1997, 
     Zavlin \& Pavlov 1998,
     Bogdanov et al. 2007 
     for J0437-4715;
     Becker et al. 2000,
     Becker \& Achenbach 2002,
     for J0030+0451;
     Zavlin 2007
     for J2124-3358;
     Zavlin 2006
     for J1024-0719),
it is suggested that the heated PC emission can be fitted
by a superposition of two thermal components.
One component has a higher temperature around
a few hundred eV with emission area of $\sim 0.3\,\mbox{km}^2$,
which is sometimes referred to as the 
\lq core' region of the heated PC.
Another component has a lower temperature around
$80$ eV with emission area of $\sim 30 \,\mbox{km}^2$,
which is sometimes referred to as the 
\lq rim' region of the heated PC.

We expect that the higher temperature region appearing in the
leading side (i.e., in the first quadrant of figure~\ref{fig:PC}) 
near the last-open field line corresponds to the observed
core region, whereas 
the lower temperature region appearing in the trailing side
(i.e., in the fourth quadrant)
partially corresponds to the rim region.
Although a detailed calculation of the heat conduction
across the NS surface may not be easy,
it is suggestive that the leading side tends to have
a higher temperature than the trailing side.
If the photon statistics will be improved in the future,
the phase-resolved spectrum of a broad peak 
in the X-ray light curve will show 
a decreasing temperature with pulsar phase. 

\acknowledgments
This work is supported by the Theoretical Institute for
Advanced Research in Astrophysics (TIARA) operated under 
Academia Sinica
and the National Science Council Excellence Projects program 
in Taiwan administered through grant number 
NSC 98-2752-M-007-006-PAE,
and the Formosa Program between National Science Council  
in Taiwan and Consejo Superior de Investigaciones Cientificas
in Spain administered through grant number 
NSC100-2923-M-007-001-MY3.

\acknowledgments

\section{Appendix}
\label{sec:app}
Derivation of equation~(\ref{eq:length}) is given below.
For simplicity, we assume a Newtonian dipole magnetic field
configuration.
We consider the plane ($x$,$z$) in which both
the rotation and magnetic axes reside.
Let us express the position of the 
cross section (point P1 in fig.~\ref{fig:sidev})
between the null surface and the last-open field line
as ($x_1$,$z_1$) in Cartesian coordinates.
Then, the radial distance of point P1 from the NS center
becomes 
\begin{equation}
  r_1 \equiv \sqrt{x_1^2+z_1^2}
      = \rlc \frac{\sin^2(\theta_{\rm null}-\alpha)}
                  {\sin^2(\theta_{\rm LC}-\alpha)},
  \label{eq:app_r1}
\end{equation}
where $\theta_{\rm null}$ specifies the null surface, and
$\theta_{\rm LC}$ the colatitudes (measured from the rotation axis)
of the point at which
the last-open field line becomes tangent to the light cylinder.
Thus, the distance of point P1 from the rotation axis becomes
\begin{equation}
  x_1 = \rlc \frac{\sin^2(\theta_{\rm null}-\alpha)}
                  {\sin^2(\theta_{\rm LC}-\alpha)}
             \sin\theta_{\rm null}
  \label{eq:app_x1}
\end{equation}
If a photon is emitted leftward from point P1,
it crosses the critical field line (fig.~\ref{fig:sidev})
before escaping from the gap at point P2 ($x_2$,$z_2$).
For a Newtonian dipole field, 
we obtain
\begin{equation}
  r_2 \equiv \sqrt{x_2^2+z_2^2}
      = \rlc \frac{\sin^2(\theta_2-\alpha)}
                  {\sin^2(\theta_{\rm null}-\alpha)},
  \label{eq:app_r2}
\end{equation}
and hence $x_2= r_2 \sin\theta_2$.
Therefore, we obtain the distance between P1 and P2,
\begin{equation}
  x_1-x_2= r_1 (\sin\theta_{\rm null}-\tan\theta_2\cos\theta_{\rm null})
 \label{eq:app_l}
\end{equation}
For photons emitted from the middle colatitude in the gap, 
we thus obtain the propagation distance below the critical field line,
\begin{equation}
  l \approx \frac{x_1-x_2}{\sqrt{2}}
    = \frac{r_1}{\sqrt{2}} 
      (\sin\theta_{\rm null}-\tan\theta_2\cos\theta_{\rm null})
 \label{eq:app_l2}
\end{equation}
where 
\begin{equation}
  \frac{\cos\alpha\tan\theta_2-\sin\alpha}
       {1+\tan^2\theta-2}
  = \frac{\cos\theta_{\rm null} \sin^4(\theta_{\rm null}-\alpha)}
         {\sin^2(\theta_{\rm LC}-\alpha)}
  \equiv \epsilon.
 \label{eq:app_eps}
\end{equation}
The factor $\sqrt{2}$ in the denominator of equation~(\ref{eq:app_l2})
is appropriate if the poloidal magnetic field 
lines are co-centric.
For a dipole field geometry on the poloidal plane, 
equation~(\ref{eq:app_l2}) overestimates $l$.
Nevertheless, the inward $\gamma$-rays propagate towards the
leading side in a three-dimensional magnetosphere
owing to aberration and propagate a longer distance than 
what is evaluated only on the poloidal plane.
Thus, equation~(\ref{eq:app_l2}) gives a good estimate.
Since $\epsilon \ll 1$, $\tan\theta_2 \approx \tan\alpha$
reduces equation~(\ref{eq:app_l2}) into
equation~(\ref{eq:length}).

\end{document}